\documentclass{ifacconf}
\usepackage{cite}
\usepackage{amsmath,amssymb,amsfonts}
\usepackage{algorithmic}
\usepackage{graphicx}
\usepackage{textcomp}
\usepackage{xcolor}
\usepackage{mathrsfs}
\usepackage{psfrag}
\usepackage{cancel}
\usepackage{comment}

\makeatletter
\let\old@ssect\@ssect 
\usepackage{natbib}
\usepackage{hyperref}
\hypersetup{colorlinks,linkcolor={blue},citecolor={blue},urlcolor={red}}
\def\@ssect#1#2#3#4#5#6{%
  \NR@gettitle{#6}
  \old@ssect{#1}{#2}{#3}{#4}{#5}{#6}
}
\makeatother

\def\mc{\mathcal}

\newtheorem{definition}{Definition}
\newtheorem{theorem}{Theorem}

\newtheorem{remark}{Remark}
\newtheorem{corollary}{Corollary}

\begin{document}
\sloppy
\begin{frontmatter}
\title{\LARGE \bf Nonlinear Negative Imaginary Systems with Switching\thanksref{footnoteinfo}}

\thanks[footnoteinfo]{This work was supported by the Australian Research Council under grant DP190102158.}

\author[First]{Kanghong Shi},
\author[First]{Ian R. Petersen},
\author[First]{Igor G. Vladimirov}

\address[First]{School of Engineering, College of Engineering and Computer Science, Australian National University, Canberra, ACT 2601, Australia.
        {\tt kanghong.shi@anu.edu.au}, {\tt ian.petersen@anu.edu.au}, {\tt igor.vladimirov@anu.edu.au}.}


\maketitle
\thispagestyle{plain}
\pagestyle{plain}

\begin{abstract}
In this paper, we extend nonlinear negative imaginary (NI) systems theory to switched systems. Switched nonlinear NI systems and switched nonlinear output strictly negative imaginary (OSNI) systems are defined. We show that the interconnection of two switched nonlinear NI systems is still switched nonlinear NI. The interconnection of a switched nonlinear NI system and a switched nonlinear OSNI system is asymptotically stable under some assumptions. This stability result is then illustrated using a numerical example.
\end{abstract}


\end{frontmatter}

\section{Introduction}
Negative imaginary (NI) systems theory was introduced in \cite{lanzon2008stability} and has attracted attention since then; see \cite{petersen2010feedback,xiong2009lossless,xiong2010negative,song2012negative,bhikkaji2011negative,wang2015robust,shi2021negative,mabrok2014generalizing}, etc. Motivated by the robust control of flexible structures \cite{preumont2018vibration,halim2001spatial,pota2002resonant}, NI systems theory uses positive position feedback control since the commonly used negative velocity feedback control may not be suitable when the velocity is not accessible to measurement. An NI system can be regarded as the cascade of a positive real (PR) system and an integrator. Typical mechanical NI systems are systems with colocated force actuators and position sensors. Compared to PR systems, which can have relative degree of zero and one (e.g., see \cite{brogliato2007dissipative}), NI systems can have relative degree of zero, one and two \cite{shi2021necessary}. Roughly speaking, a square real-rational proper transfer matrix $F(s)$ is said to be NI if it is stable and $j(F(j\omega)-F(j\omega)^*)\geq 0$ for all $\omega \geq 0$. NI systems theory has been applied in many fields including nano-positioning control \cite{mabrok2013spectral,das2014mimo,das2014resonant,das2015multivariable} and the control of lightly damped structures \cite{cai2010stability,rahman2015design,bhikkaji2011negative}, etc.

Motivated by the fact that many control systems are nonlinear, NI systems theory was extended to nonlinear systems in \cite{ghallab2018extending}. The definition of nonlinear NI systems was then extended in \cite{shi2021robust} to include multi-input multi-output (MIMO) systems. The definition was extended again in \cite{shi2022output} to allow for systems with free body motion. More precisely, a system is said to be nonlinear NI if there exists a positive semidefinite function $V(x)$ such that its time derivative satisfies $\dot V(x)\leq u^T\dot y$, where $x,u,y$ are the state, input and output of the system, respectively. Under some assumptions, the closed-loop interconnection of a nonlinear NI system and a nonlinear output strict negative imaginary (OSNI) system (e.g., see \cite{bhowmick2017lti,bhowmickoutput} for linear OSNI systems) is asymptotically stable.

Switched systems, which form an important class of hybrid systems, have drawn much attention because many real-world dynamical systems are of hybrid nature; e.g., see \cite{liberzon2003switching,brockett1993hybrid}. Such systems are organized as continuous-time systems with discrete-time switching events and arise in many control applications \cite{decarlo2000perspectives,narendra2003adaptive,sun2006switched}. Besides this, the study of switched systems is also motivated by the improvement of control performance that can be achieved using switched controllers \cite{kolmanovsky2006multi,skafidas1999stability,deenen2017hybrid,heertjes2019robust}.

Stability problems for switched systems have been investigated in many books and papers for general systems (e.g., see \cite{liberzon2003switching,goebel2009hybrid,liberzon1999basic} and also specifically for dissipative systems including passive systems \cite{zhao2008dissipativity,zhao2008passivity,chen2005passivity}. However, control problems for switched nonlinear NI systems have never been investigated. As nonlinear NI systems theory complements the passivity theory by allowing systems to have relative degree two, and such systems arise in many control applications (see \cite{petersen2016negative} and references therein), it is worth investigating the class of switched nonlinear NI systems and the corresponding stability problems.

The present paper extends the nonlinear NI systems theory to switched systems. We provide definitions of switched nonlinear NI and OSNI systems. A switched nonlinear system is said to be NI if all the subsystems being switched to are NI, and in addition, the storage functions for the subsystems in the switching sequence are nonincreasing. Switched nonlinear OSNI systems are defined accordingly. It is shown that with external inputs and outputs considered, the closed-loop interconnection of two switched nonlinear NI (OSNI) systems is also switched NI (OSNI). Under zero input, the interconnection of two switched nonlinear NI systems are stable in the sense of Lyapunov. The main result of this paper is that the closed-loop interconnection of a switched nonlinear NI system and a switched nonlinear OSNI system is asymptotically stable under some assumptions. This stability result is illustrated by an example where a nonlinear mass-spring-damper system is asymptotically stabilized by a certain switched nonlinear NI controller.

This rest of the paper is organized as follows: Section \ref{sec:preliminaries} provides the definitions for nonlinear NI and OSNI systems without switching. Switched nonlinear NI and OSNI systems are defined in Section \ref{sec:definition} and the NI property is investigated for the interconnection of two switched nonlinear NI systems with external inputs. The main results of this paper are presented in Section \ref{sec:stability}, where the stability of the interconnection of switched nonlinear NI systems are investigated. Section \ref{sec:example} illustrates the stability results using a control example where a nonlinear mass-spring-damper system is stabilized by a switched nonlinear NI controller. Section \ref{sec:conclusion} concludes the paper.  

\textbf{Notation}: The notation in this paper is standard. $\mathbb R$ denotes the set of real numbers. $\mathbb N$ denotes the set of nonnegative integer numbers. $\mathbb R^{m\times n}$ denotes the space of real matrices of dimension $m\times n$. $A^T$ and $A^*$ denote the transpose and complex conjugate transpose of a matrix $A$, respectively. $\overline{(\cdot)}$ denotes a constant value for a given vector or scalar signal. $\|\cdot \|$ denotes the Euclidean norm of a vector. $C^k$ denotes the class of $k$ times continuously differentiable functions.

\section{Preliminaries}\label{sec:preliminaries}
Consider the following general nonlinear system:
\begin{subequations}\label{eq:general nonlinear system}
	\begin{align}
    \dot x(t)=&\ f(x(t),u(t)),\label{eq:state equation of nonlinear NI}\\
    y(t)=&\ h(x(t)),
    \label{eq:output equation of nonlinear NI}
\end{align}
\end{subequations}
where $x(t)\in \mathbb R^{n}$ is the state, $u(t)\in \mathbb R^p$ is a locally integrable input, and $y(t)\in \mathbb R^p$ is the output, $f:\mathbb R^n\times \mathbb R^p \to \mathbb R^n$ is a Lipschitz continuous function and $h:\mathbb R^n \to \mathbb R^p$ is a class $C^1$ function.
\begin{definition}\label{def:nonlinear NI}\cite{ghallab2018extending,shi2021robust,shi2022output}
The system (\ref{eq:general nonlinear system}) is said to be a nonlinear NI system if there exists a positive semidefinite storage function $V:\mathbb R^n\to \mathbb R$ of class $C^1$ such that for any locally integrable input $u$ and any solution $x$ to (\ref{eq:general nonlinear system}),
\begin{equation}\label{eq:NI MIMO definition inequality}
    \dot V(x(t))\leq u(t)^T\dot y(t),
\end{equation}
for all $t\geq 0$.
\end{definition}

\begin{definition}\label{def:nonlinear OSNI}
The system (\ref{eq:general nonlinear system}) is said to be a nonlinear output strictly negative imaginary (OSNI) system if there exists a positive semidefinite storage function $V:\mathbb R^n\to\mathbb R$ of class $C^1$ and a scalar $\epsilon>0$ such that for any locally integrable input and any solution to (\ref{eq:general nonlinear system}),
\begin{equation}\label{eq:dissipativity of OSNI}
    \dot V(x(t))\leq u(t)^T\dot y(t) -\epsilon \left\|\dot y(t)\right\|^2,
\end{equation}
for all $t\geq 0$. In this case, we also say that system (\ref{eq:general nonlinear system}) is nonlinear OSNI with degree of output strictness $\epsilon$.
\end{definition}
\section{Switched nonlinear NI systems}\label{sec:definition}
In this section, we provide definitions of nonlinear NI and OSNI properties for switched systems. We show that the positive feedback interconnection of two switched nonlinear NI systems, with external inputs and outputs considered, is also a switched nonlinear NI system. A similar result holds for switched nonlinear OSNI systems.

Consider a continuous time switched system of the following form (e.g., see \cite{lin2009stability,liberzon1999basic,liberzon2003switching})
\begin{subequations}\label{eq:switched system}
	\begin{align}
		\dot x =& f_\sigma(x,u),\\
		y =& h(x),
	\end{align}
\end{subequations}
where $x(t)\in \mathbb R^{n}$ is the absolutely continuous state, $u(t)\in \mathbb R^p$ is a locally integrable input, and $y(t)\in \mathbb R^p$ is the output. Here, $h:\mathbb R^n \to \mathbb R^p$ is a class $C^1$ function and $\{f_i:\mathbb R^n\times \mathbb R^p \to \mathbb R^n, i\in \mc I\}$ is a family of Lipschitz continuous functions parametrized by an index set $\mc I=\{1,2,\cdots,N\}$. The \emph{switching signal} $\sigma:[0,\infty)\to \mc I$ is a piecewise constant function of time. The index $i=\sigma(t)$ is called the active mode at time $t$. We assume that $\sigma$ is right continuous everywhere; i.e., $\sigma(t) = lim_{\tau\to t^+}\sigma(\tau)$ for each $t\geq 0$. In general, the value of $\sigma$ at time $t$ can depend on $t$, $x(t)$ or both, or be generated using hybrid feedback with memory in the loop. Since the state $x(t)$ of (\ref{eq:switched system}) is an absolutely continuous function of time, it does not jump at switching instants. We assume that the system switches only a finite number of times on any finite time interval; i.e., there is no chattering. Then, we can denote by $\{(i_0,t_0), (i_1,t_1), \cdots, (i_k,t_k), \cdots\}$, $(k\in \mathbb N)$ the switching index sequence. Here, $t_0$ is the initial time and $\mathbb N$ denotes the set of nonnegative integers. This means that at time $t_k$, the state equation of the system becomes $\dot x = f_{i_k}(x,u)$. Note that the switched system has a common state, a common input and a common output, while the dynamic equation of the system switches to another map $f$ at every switching instant.

Now, we define switched nonlinear NI systems as follows.
\begin{definition}\label{def:switched NI}
	The switched nonlinear system (\ref{eq:switched system}) is said to be a switched nonlinear NI system if the following two conditions hold:
	
	1. For all $i\in \mc I$, there exist positive semidefinite functions $V_{i}(x)$ such that for any input $u$ and any solution $x$ to (\ref{eq:switched system}) with $\sigma = i$,
	\begin{equation}\label{eq:switched NI ineq}
	\frac{\partial V_{i}(x)}{\partial x}f_{i}(x,u) \leq u^T\frac{\partial h(x)}{\partial x}f_{i}(x,u).
	\end{equation}

2. For all $k\in \mathbb N$, $V_{i_{k+1}}(x)\leq V_{i_{k}}(x)$ for all $x\in\mathbb R^n$.
\end{definition}

Similarly, we define switched nonlinear OSNI systems as follows.
\begin{definition}\label{def:switched OSNI}
	The switched nonlinear system (\ref{eq:switched system}) is said to be a switched nonlinear OSNI system if the following two conditions hold:
	
	1. For all $i\in \mc I$, there exist positive semidefinite functions $V_{i}(x)$ such that for any input $u \in \mathbb R^p$ and any solution $x\in \mathbb R^n$ to (\ref{eq:switched system}) with $\sigma = i$,
	\begin{equation}\label{eq:switched OSNI ineq}
		\frac{\partial V_{i}(x)}{\partial x}f_{i}(x,u) \leq u^T\frac{\partial h(x)}{\partial x}f_{i}(x,u)-\epsilon\|\frac{\partial h(x)}{\partial x}f_{i}(x,u)\|^2.
	\end{equation}

2. For all $k\in \mathbb N$, $V_{i_{k+1}}(x)\leq V_{i_{k}}(x)$ for all $x\in \mathbb R^n$.
\end{definition}

\begin{remark}
	Definitions \ref{def:switched NI} and \ref{def:switched OSNI} state that a system is switched nonlinear NI (OSNI) if every subsystem satisfies the regular nonlinear NI (OSNI) system definition as given in Definitions \ref{def:nonlinear NI} and \ref{def:nonlinear OSNI} and in addition, the system switches in a pattern such that the storage functions of the subsystems in the switching index sequence are nonincreasing. Note that inequalities (\ref{eq:switched NI ineq}) and (\ref{eq:switched OSNI ineq}) correspond to inequalities (\ref{eq:NI MIMO definition inequality}) and (\ref{eq:dissipativity of OSNI}), respectively. The time derivatives are expressed using the directional derivatives in order to specify the NI and OSNI inequalities for the subsystem that corresponds to the current switching index $i_k$.
\end{remark}

\begin{figure}[h!]
\centering
\psfrag{H_1}{$H_1$}
\psfrag{in_0}{$u$}
\psfrag{y_1}{$y$}
\psfrag{in_2}{$\tilde e$}
\psfrag{y_2}{$\tilde y$}
\psfrag{in_1}{$e$}
\psfrag{H_2}{$H_2$}
\psfrag{in_3}{$\tilde u$}
\hspace{0.5cm}\includegraphics[width=7.5cm]{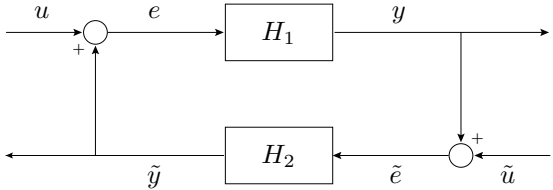}
\caption{Closed-loop interconnection of switched nonlinear systems $H_1$ and $H_2$ with external inputs $u$, $\tilde u$ and external outputs $y$, $\tilde y$.}
\label{fig:closed-loop with inputs}
\end{figure}

We investigate in the following the switched nonlinear NI (OSNI) property for the interconnection of two switched nonlinear NI (OSNI) systems with external inputs and outputs, as shown in Fig.~\ref{fig:closed-loop with inputs}. In what follows, the symbol ``tilde", that is $\tilde \cdot$, indicates the variables associated with the system $H_2$.

Suppose the switched systems $H_1$ and $H_2$ in the closed-loop interconnection shown in Fig.~\ref{fig:closed-loop with inputs} have, without loss of generality, synchronized switching instants. This also applies to the interconnection shown below in Fig.~\ref{fig:closed-loop interconnection}.

\begin{theorem}\label{thm:switched NI}
	Suppose the systems $H_1$ and $H_2$ in the positive feedback interconnection in Fig.~\ref{fig:closed-loop with inputs} are switched nonlinear NI (OSNI) systems of the form (\ref{eq:switched system}) with states $x\in \mathbb R^n, \tilde x\in \mathbb R^{\tilde n}$, inputs $e, \tilde e\in \mathbb R^p$ and outputs $y,\tilde y\in\mathbb R^p$, respectively. Also, suppose that for all $i_k \in \mc I$,
	\begin{equation}\label{eq:psd W in switched NI thm}
V_{i_k}(x)+\tilde V_{i_k}(\tilde x)-h(x)^T\tilde h(\tilde x)\geq 0,
	\end{equation}
for all $x\in \mathbb R^n, \tilde x \in \mathbb R^{\tilde n}$, where $V_{i_k}(x)$ and $\tilde V_{i_k}(\tilde x)$ are positive semidefinite storage functions that satisfy Definition \ref{def:switched NI} (Definition \ref{def:switched OSNI}) for the systems $H_1$ and $H_2$, respectively. Then the closed-loop system with input $\widehat u=\left[\begin{matrix}
		u^T & \tilde u^T
	\end{matrix}\right]^T$ and output $\widehat y=\left[\begin{matrix}
		y^T & \tilde y^T
	\end{matrix}\right]^T$ is also a switched nonlinear NI (OSNI) system.
\end{theorem}
\begin{pf}
	We first prove the OSNI part of the theorem. Suppose $H_1$ and $H_2$ are switched nonlinear OSNI systems. Then according to Definition \ref{def:switched OSNI}, there exists positive semidefinite functions $V_{i_k}(x)$ and $\tilde V_{i_k}(\tilde x)$ for every switching index $i_k\in \mc I$ such that
	\begin{align}
		\frac{\partial V_{i_k}(x)}{\partial x}f_{i_k}(x,e)\leq &e^T\frac{\partial h(x)}{\partial x}f_{i_k}(x,e)-\epsilon\|\frac{\partial h(x)}{\partial x}f_{i_k}(x,e)\|^2;\notag\\
		\frac{\partial {\tilde V}_{i_k}(\tilde x)}{\partial {\tilde x}}{\tilde f}_{i_k}(\tilde x,\tilde e)\leq &\tilde e^T\frac{\partial {\tilde h}(\tilde x)}{\partial {\tilde x}}\tilde f_{i_k}(\tilde x,\tilde e)-\tilde \epsilon\|\frac{\partial {\tilde h}(\tilde x)}{\partial {\tilde x}}\tilde f_{i_k}(\tilde x,\tilde e)\|^2,\notag
	\end{align}
	for all $e$, $\tilde e$ and $x$, $\tilde x$. In order to satisfy the conditions in Definition \ref{def:switched OSNI}, we construct the storage function for the closed-loop system to be
	\begin{equation}\label{eq:storage function for switched NI property}
		W_{i_k}(x,\tilde x) = V_{i_k}(x)+\tilde V_{i_k}(\tilde x)-h(x)^T\tilde h(\tilde x)
	\end{equation}
for the switching index $i_k\in \mc I$. According to (\ref{eq:psd W in switched NI thm}), the function $W_{i_k}(x,\tilde x)$ is positive semidefinite. For condition 1 in Definition \ref{def:switched OSNI}, we have that
\begin{align}
	\dot W_{i_k}(x,\tilde x)=&\ \dot V_{i_k}+\dot {\tilde V}_{i_k}-\dot h(x)^T\tilde h(\tilde x)-h(x)^T\dot {\tilde h}(\tilde x)\notag\\
 \leq &\ e^T\dot y-\epsilon\|\dot y\|^2+\tilde e^T\dot {\tilde y}-\tilde \epsilon\|\dot {\tilde y}\|^2-\dot y^T\tilde y-y^T\dot {\tilde y}\notag\\
 = &\ (u+\tilde y)^T\dot y-\epsilon\|\dot y\|^2+(\tilde u+y)^T\dot {\tilde y}-\tilde \epsilon\|\dot {\tilde y}\|^2\notag\\
 &-\dot y^T\tilde y-y^T\dot {\tilde y}\notag\\
 =&\ u^T\dot y+ \tilde u^T\dot {\tilde y} -\epsilon\|\dot y\|^2-\tilde \epsilon\|\dot {\tilde y}\|^2\notag\\
 =& \widehat u^T \dot {\widehat y} - \epsilon_{min}\|\dot {\widehat y}\|^2,\label{eq:switched NI ineq for NI property}
\end{align}
where the system settings $e = u+\tilde y$ and $\tilde e = \tilde u+y$ as shown in Fig.~\ref{fig:closed-loop with inputs} are also used. Here, $\epsilon_{min} = \min \{\epsilon,\tilde \epsilon\}>0$ is the level of output strictness for the interconnection of $H_1$ and $H_2$. For Condition 2 in Definition \ref{def:switched OSNI}, we have that
\begin{equation}\label{eq:W jumps down}
	W_{i_{k+1}}(x,\tilde x) \leq W_{i_{k}}(x,\tilde x)
\end{equation}
for all $k\in \mathbb N$ because $V_{i_{k+1}}(x)\leq V_{i_{k}}(x)$, $\tilde V_{i_{k+1}}(\tilde x)\leq \tilde V_{i_{k}}(\tilde x)$, and $h(x)$ and $\tilde h(\tilde x)$ are continuous. Therefore, since Conditions 1 and 2 in Definition \ref{def:switched OSNI} are both satisfied, then the system shown in Fig.~\ref{fig:closed-loop with inputs} is a switched nonlinear OSNI system. In the case when $H_1$ and $H_2$ are both switched nonlinear NI, the proof of the switched nonlinear NI property for their interconnection follows from a similar analysis in the limiting case of $\epsilon=\tilde \epsilon=0$. \hfill $\blacksquare$
\end{pf}

\section{Stability for the interconnection of switched nonlinear NI systems}\label{sec:stability}

In this section, we show that the interconnection of two switched nonlinear NI systems without external inputs is Lyapunov stable. Also, under some assumptions, the interconnection of a switched nonlinear NI system and a switched nonlinear OSNI system is asymptotically stable.

\begin{figure}[h!]
\centering
\psfrag{H_1}{$H_1$}
\psfrag{in_0}{$r=0$}
\psfrag{y_1}{$y$}
\psfrag{in_2}{\vspace{-0.5cm}$\tilde u$}
\psfrag{y_2}{\vspace{-0.5cm}$\tilde y$}
\psfrag{in_1}{$u$}
\psfrag{H_2}{$H_2$}
\hspace{0.5cm}\includegraphics[width=9cm]{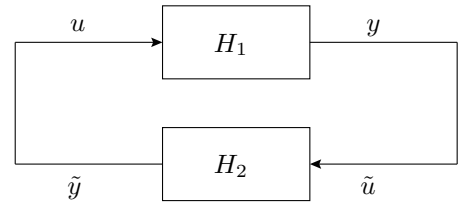}
\caption{Closed-loop interconnection of switched nonlinear systems $H_1$ and $H_2$.}
\label{fig:closed-loop interconnection}
\end{figure}

\begin{theorem}
	Suppose the systems $H_1$ and $H_2$ in the interconnection in Fig.~\ref{fig:closed-loop interconnection} are switched nonlinear NI systems of the form (\ref{eq:switched system}) with states $x\in \mathbb R^n, \tilde x\in \mathbb R^{\tilde n}$, inputs $u, \tilde u\in \mathbb R^p$ and outputs $y,\tilde y\in\mathbb R^p$, respectively. Also, suppose for all $i_k \in \mc I$, the function
	\begin{equation}\label{eq:pd assumption for W}
V_{i_k}(x)+\tilde V_{i_k}(\tilde x)-h(x)^T\tilde h(\tilde x)
	\end{equation}
is positive definite for all $x\in \mathbb R^n, \tilde x \in \mathbb R^{\tilde n}$, where $V_{i_k}(x)$ and $\tilde V_{i_k}(\tilde x)$ are storage functions satisfying Definition \ref{def:switched NI} for systems $H_1$ and $H_2$, respectively. Then the interconnection of $H_1$ and $H_2$ shown in Fig.~\ref{fig:closed-loop interconnection} is Lyapunov stable.
\end{theorem}
\begin{pf}
	According to Theorem \ref{thm:switched NI}, the interconnection of two switched nonlinear NI systems is also a switched nonlinear NI system. The function in (\ref{eq:storage function for switched NI property}) satisfies $\dot W_{i_k}(x,\tilde x)\leq 0$ when there is no external input, according to (\ref{eq:switched NI ineq for NI property}). Assuming the function given by (\ref{eq:pd assumption for W}) to be positive definite ensures that $W_{i_k}(x,\tilde x)$ is a Lyapunov function for every switching index $i_k$. Also, since $W_{i_k}(x,\tilde x)$ satisfies (\ref{eq:W jumps down}), then we conclude that $W_{\sigma(t_2)}(x,\tilde x)\leq W_{\sigma(t_1)}(x,\tilde x)$ for any $t_2>t_1 \geq 0$. Therefore, the interconnection of $H_1$ and $H_2$ is Lyapunov stable. \hfill $\blacksquare$
\end{pf}

It is natural to ask whether the interconnection of $H_1$ and $H_2$ shown in Fig.~\ref{fig:closed-loop interconnection} is asymptotically stable if $H_1$ and $H_2$ are both switched nonlinear OSNI systems. Indeed, the interconnection in Fig.~\ref{fig:closed-loop interconnection} is asymptotically stable if at least one of the systems $H_1$ and $H_2$ is OSNI. This is shown in Theorem \ref{theorem:stability}, where several additional assumptions on the systems $H_1$ and $H_2$ are needed. For simplicity of notation, we omit the ``tilde" in the following two assumptions.

\textbf{Assumption I}: Over any time interval $[t_a,t_b]$ with $t_b>t_a$, $h(x(t))$ remains constant if and only if $x(t)$ remains constant; that is $\dot h(x(t))\equiv 0\iff \dot x(t)\equiv 0$. Moreover, $h(x(t))\equiv 0 \iff x(t)\equiv 0$.

\textbf{Assumption II}: Over any time interval $[t_a,t_b]$ with $t_b>t_a$, $x(t)$ remains constant only if $u(t)$ remains constant; that is $x(t)\equiv \bar x \implies u(t)\equiv\bar u$. Moreover, $x(t)\equiv 0\implies u(t)\equiv 0$.

\begin{figure}[h!]
\centering
\psfrag{in_1}{$u$}
\psfrag{y_1}{$y$}
\psfrag{in_2}{\hspace{0.3cm}$\tilde u$}
\psfrag{y_2}{$\tilde y$}
\psfrag{H_1}{$H_1$}
\psfrag{H_2}{$H_2$}
\includegraphics[width=8cm]{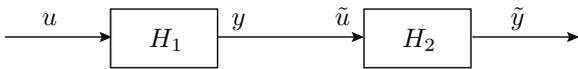}
\caption{Cascade interconnection of systems $H_1$ and $H_2$.}
\label{fig:open-loop interconnection of single NI and OSNI}
\end{figure}

Also, for the cascade interconnection of the systems $H_1$ and $H_2$ shown in Fig.~\ref{fig:open-loop interconnection of single NI and OSNI}, suppose:

\textbf{Assumption III}: Given any nonzero constant input $\bar u$ for the system $H_1$, we obtain a corresponding output $y(t)$. Set the input of the system $H_2$ to be $\tilde u(t)\equiv y(t)$ in the cascade interconnection in Fig.~\ref{fig:open-loop interconnection of single NI and OSNI}. If the corresponding output of the system $H_2$ is constant; that is $\tilde y(t)\equiv \bar {\tilde y}$, then we have
\begin{equation}\label{eq:DC gain condition}
    \bar u\neq \bar {\tilde y}.
\end{equation}
\vspace{-0.5cm}

\begin{remark}	
Assumptions I and II are nonlinear generalizations of standard properties for linear systems. Assumption I is an observability assumption. Assumption II requires all inputs to be involved in affecting the system dynamics, which can be supposed without loss of generality. In the case of linear systems, Assumption II requires the matrix $B$ in the realisation $(A,B,C,D)$ of the linear system to have full column rank.

Assumption III requires the steady-state input and output of the cascade interconnection of $H_1$ and $H_2$ in Fig.~\ref{fig:open-loop interconnection of single NI and OSNI} to be different for nonzero inputs. This assumption and the positive definiteness condition in Theorem \ref{theorem:stability} form a nonlinear counterpart of the DC gain condition used in linear NI systems stability results (see for example \cite{lanzon2008stability,petersen2010feedback}).
\end{remark}

\begin{theorem}\label{theorem:stability}
	Suppose a switched nonlinear NI system $H_1$ and a switched nonlinear OSNI system $H_2$, both of the form (\ref{eq:switched system}) have states $x\in \mathbb R^n, \tilde x\in \mathbb R^{\tilde n}$, inputs $u,\tilde u\in \mathbb R^p$ and outputs $y,\tilde y \in \mathbb R^p$, respectively, and are in the positive feedback interconnection in Fig.~\ref{fig:closed-loop interconnection}. Also, suppose Assumptions I-III are satisfied and for all $i_k\in \mc I$, the storage function
\begin{align}
    W_{i_k}(x,\tilde x):=& \ V_{i_k}(x)+\tilde V_{i_k}(\tilde x)-h(x)^T\tilde h(\tilde x)\label{eq:W},
\end{align}
is positive definite for all $x\in \mathbb R^n, \tilde x\in \mathbb R^{\tilde n}$, where $V_{i_k}(x)$ and $\tilde V_{i_k}(\tilde x)$ are positive semidefinite storage functions satisfying Definition \ref{def:switched NI} for the system $H_1$ and Definition \ref{def:switched OSNI} for the system $H_2$, respectively. Then the closed-loop interconnection of the systems $H_1$ and $H_2$ in Fig.~\ref{fig:closed-loop interconnection} is asymptotically stable.
\end{theorem}
\begin{pf}
	We apply Lyapunov's direct method. The storage function defined in (\ref{eq:W}) is positive definite for all switching indices $i_k \in \mc I$. Therefore, the storage function for the closed-loop system $W_\sigma(x,\tilde x)$ is positive definite. We need to prove that $W_\sigma(x,\tilde x)$ keeps decreasing until it reaches zero. We break the proof into two parts. First we show that for $t\in (t_k,t_{k+1})$ where $k \in \mathbb N$, the function $W_{i_k}(x,\tilde x)$ keeps decreasing unless $x=0$. Taking the time derivative of $W_{i_k}(x,\tilde x)$, we have
	\begin{align}
		\dot W_{i_k}&(x,\tilde x)\notag\\
		=&\ \dot V_{i_k}(x)+\dot {\tilde V}_{i_k}(\tilde x)-\dot h(x)^T\tilde h(\tilde x)-h(x)^T\dot {\tilde h}(\tilde x)\notag\\
	=&\ \frac{\partial V_{i_k}(x)}{\partial x}f_{i_k}(x,u)+\frac{\partial {\tilde V}_{i_k}(\tilde x)}{\partial \tilde x}f_{i_k}(\tilde x,\tilde u)\notag\\
	&-\left(\frac{\partial h(x)}{\partial x}f_{i_k}(x,u)\right)^T \tilde h(\tilde x)-h(x)^T\frac{\partial \tilde h(\tilde x)}{\partial \tilde x}\tilde f(\tilde x,\tilde u)\notag\\
	\leq &\ u^T\frac{\partial h(x)}{\partial x}f_{i_k}(x,u)+\tilde u^T\frac{\partial \tilde h(\tilde x)}{\partial \tilde x}\tilde f_{i_k}(\tilde x,\tilde u)\notag\\
	&-\tilde \epsilon \left\|\frac{\partial \tilde h(\tilde x)}{\partial \tilde x}\tilde f_{i_k}(\tilde x,\tilde u)\right\|^2-\left(\frac{\partial h(x)}{\partial x}f_{i_k}(x,u)\right)^T \tilde h(\tilde x)\notag\\
	&-h(x)^T\frac{\partial \tilde h(\tilde x)}{\partial \tilde x}\tilde f(\tilde x,\tilde u)\notag\\
	=& -\tilde \epsilon \left\|\frac{\partial \tilde h(\tilde x)}{\partial \tilde x}\tilde f_{i_k}(\tilde x,\tilde u)\right\|^2\leq  0,\label{eq:W ineq}
	\end{align}
where the equalities also use the closed-loop setting that $u = \tilde y=\tilde h(\tilde x)$ and $\tilde u=y=h(x)$. The inequality (\ref{eq:W ineq}) implies that $\dot W_{i_k}(x,\tilde x)$ is non-increasing. Now we apply LaSalle's invariance principle. According to (\ref{eq:W ineq}), $\dot W_{i_k}(x,\tilde x)$ remains zero only if $\frac{\partial \tilde h(\tilde x)}{\partial \tilde x}\tilde f_{i_k}(\tilde x,\tilde u)$ remains zero. This implies that $\tilde x$ remains constant according to Assumption I. According to Assumption II, $\tilde x$ remaining constant implies that $\tilde u$ remains constant. Considering the setting $\tilde u=y$ as shown in Fig.~\ref{fig:closed-loop interconnection}, we have that $y$ also remains constant. This implies that both $x$ and $u$ remain constant according to Assumptions I and II. Therefore, the system is in a steady state. In this case, the inputs $u(t)$, $\tilde u(t)$, states $x(t)$, $\tilde x(t)$ and outputs $y(t)$, $\tilde y(t)$ are all constant vectors, which we denote by $\bar u$, $\bar {\tilde u}$, $\bar x$, $\bar {\tilde x}$, $\bar y$, $\bar {\tilde y}$, respectively. According to Assumption III, if $\bar u \neq 0$, then $\bar u \neq \bar {\tilde y}$, which contradicts $\bar u=\bar {\tilde y}$. This implies that (\ref{eq:W ineq}) can hold only if $\bar u=\bar {\tilde y}=0$. According to Assumptions I and II, we have that $\bar {\tilde y}=0 \implies \bar {\tilde x}=0 \implies \bar {\tilde u}=\bar y=0 \implies \bar x =0$. This means that the system is already in equilibrium. Otherwise, $W_{i_k}(x,\tilde x)$ keeps decreasing until it reaches zero. This completes the first part of the proof.

We also need to prove that $W_\sigma(x,\tilde x)$ does not increase at all switching instants $t_k$. At a switching instant $t_{k+1}$, the switching index changes from $i_k$ to $i_{k+1}$, which leads to a jump of $W_\sigma(x,\tilde x)$ from $W_{i_k}(x(t_{k+1}),\tilde x(t_{k+1}))$ to $W_{i_{k+1}}(x(t_{k+1}),\tilde x(t_{k+1}))$. Note that there is no jump in the states $x$ and $\tilde x$. Therefore, using (\ref{eq:W}), we have that
	\begin{align*}
		W_{i_{k+1}}&(x(t_{k+1}),\tilde x(t_{k+1}))- W_{i_{k}}(x(t_{k+1}),\tilde x(t_{k+1}))\notag\\
		=&V_{i_{k+1}}(x(t_{k+1}))+\tilde V_{i_{k+1}}(\tilde x(t_{k+1}))\notag\\
		&-h(x(t_{k+1}))^T\tilde h(\tilde x(t_{k+1}))-V_{i_{k}}(x(t_{k+1}))\notag\\
		&-\tilde V_{i_{k}}(\tilde x(t_{k+1}))+h(x(t_{k+1}))^T\tilde h(\tilde x(t_{k+1}))\notag\\
		=&V_{i_{k+1}}(x(t_{k+1}))-V_{i_{k}}(x(t_{k+1}))+\tilde V_{i_{k+1}}(\tilde x(t_{k+1}))\notag\\
		&-\tilde V_{i_{k}}(\tilde x(t_{k+1}))\leq 0,
	\end{align*}
according to Condition 2 in Definitions \ref{def:switched NI} and \ref{def:switched OSNI}. Therefore, we conclude that the storage function $W_\sigma(x,\tilde x)$ keeps decreasing monotonically except for $(x,\tilde x)=(0,0)$. Since $W_\sigma(x,\tilde x)\geq 0$, then $W_\sigma(x,\tilde x)$ is bounded from below and hence $W_\sigma(x,\tilde x)$ has a limit $c$ as $t\to \infty$. We prove by contradiction that $c=0$. Suppose $c> 0$, then it implies that $\left[\begin{matrix}
	x(t)^T & \tilde x(t)^T
\end{matrix}\right]^T\neq 0$ for all $t\geq 0$. In this case, $(x,\tilde x)$ converges to a limit set $\mc S$ such that $(x,\tilde x)\in \mc S$ implies $\dot W_\sigma(x,\tilde x)= 0$. However, as is proved above, $\dot W_\sigma(x,\tilde x)$ remains zero only at $(x,\tilde x)=(0,0)$. This contradicts $\left[\begin{matrix}
	x(t)^T & \tilde x(t)^T
\end{matrix}\right]^T\neq 0$. Therefore, $c = 0$ and the system is asymptotically stable. \hfill $\blacksquare$
\end{pf}

\begin{remark}
	A special case of Theorem \ref{theorem:stability} is when there exists a common storage function for all switching indices for the system $H_1$, and respectively the system $H_2$; that is $V_{i}(\cdot) = V_{j}(\cdot)$ for all $i,j\in \mc I$ and $\tilde V_{i}(\cdot) = \tilde V_{j}(\cdot)$ for all $i,j\in \tilde {\mc I}$. Although this special case is included in Theorem \ref{theorem:stability}, it is important for switched systems because it allows a switched system with a finite index set $\mc I$ to have infinite and arbitrary switching sequence. Hence, we provide a stability result particularly for this case.
\end{remark}

\begin{corollary}\label{corollary:common Lyapunov function}
	For the system setting in Theorem \ref{theorem:stability},	suppose $V_{i}(\cdot) = V_{j}(\cdot)$ for all $i,j\in \mc I$ and $\tilde V_{i}(\cdot) = \tilde V_{j}(\cdot)$ for all $i,j\in \tilde {\mc I}$. Also, suppose Assumptions I-III are satisfied, and the storage function, given by
\begin{equation*}
    W(x,\tilde x):= \ V(x)+\tilde V(\tilde x)-h(x)^T\tilde h(\tilde x),
\end{equation*}
is positive definite for all $x\in \mathbb R^n, \tilde x\in \mathbb R^{\tilde n}$, where $V(x)$ and $\tilde V(\tilde x)$ are the common positive semidefinite storage functions satisfying Definition \ref{def:switched NI} for the system $H_1$ and Definition \ref{def:switched OSNI} for the system $H_2$, respectively. Then the closed-loop interconnection for the systems $H_1$ and $H_2$ in Fig.~\ref{fig:closed-loop interconnection} is asymptotically stable.
\end{corollary}
\begin{pf}
	Since Corollary \ref{corollary:common Lyapunov function} considers a special yet important case of Theorem \ref{theorem:stability}, the proof follows directly from that of Theorem \ref{theorem:stability}.\hfill $\blacksquare$
\end{pf}

\section{Illustrative Example}\label{sec:example}
In this section, we provide a numerical example demonstrating the stability of the interconnection for a switched nonlinear NI system and a switched nonlinear OSNI system of Section \ref{sec:stability}. Here, we apply a state dependent switched nonlinear NI system called hybrid integrator-gain system (HIGS) as a controller to a mass-spring-damper system with a nonlinear spring. HIGS were developed in \cite{deenen2017hybrid} to improve low-frequency disturbance rejection behaviour and to reduce high-frequency amplification and overshoot \cite{heertjes2019robust}. This is realized by a switching mechanism between an integrator mode and a gain mode. The system model of a HIGS is given in the following (see \cite{deenen2017hybrid}):
\begin{equation}\label{eq:HIGS}
		\mathcal{H}:
		\begin{cases}
			\dot{x}_h(t) = \omega_h e(t), & \text{if}\, (e(t),u(t),\dot{e}(t)) \in \mathcal{F}_1\\
			x_h(t) = k_he(t), & \text{if}\, (e(t),u(t),\dot{e}(t)) \in \mathcal{F}_2\\
			u(t) = x_h(t)
		\end{cases}
	\end{equation}
where $x_h(t),e(t),u(t) \in \mathbb{R}$ denote the HIGS state, input and output, respectively. For convenience, we omit the time arguments in what follows. Here, $\dot{e}$ is the time derivative of the input $e$ which is assumed to be continuous and piecewise differentiable. The parameters $\omega_h \in [0,\infty)$ and $k_h \in (0, \infty)$ are the integrator frequency and gain value, respectively. The sets $\mathcal{F}_1,\mathcal{F}_2 \in \mathbb{R}^3$ determine the HIGS modes of operation; that is, the integrator and gain modes, respectively. By construction, $\mathcal{F} = \mathcal{F}_1 \cup \mathcal{F}_2$ represents the sector $[0, k_h]$ as (see \cite{deenen2017hybrid})
\begin{equation*}\label{eq:subspace_F}
	\mathcal{F} = \{ (e,u,\dot{e}) \in \mathbb{R}^3 |\, eu \geq \frac{u^2}{k_h}\},
\end{equation*}
and $\mathcal{F}_1$ and $\mathcal{F}_2$ are defined as
	\begin{align*}
	\mathcal{F}_1& := \mathcal{F} \setminus \mathcal{F}_2;\\
	\mathcal{F}_2& := \{(e,u,\dot{e}) \in \mathbb{R}^3 | u = k_he\quad \text{and}\quad  \omega_he^2 > k_he\dot{e}\}.
\end{align*}
The HIGS primarily operates in the integrator mode. It will be switched to the gain mode once the system dynamics reach the boundary of $\mc F$ and tend to leave $\mc F$ if it stays in the integrator mode.

As is shown in \cite{shi2022negative}, the two modes of the HIGS (\ref{eq:HIGS}) satisfy the inequality (\ref{eq:switched NI ineq}) with a common Lyapunov function
\begin{equation*}
	V(x) = \frac{1}{2k_h}x_h^2.
\end{equation*}
We consider a nonlinear OSNI mass-spring-damper system of the following  form to be the plant
\begin{align}
	H_1: \quad \dot x_1 =&\ x_2,\notag\\
	\dot x_2 =& -x_1^3-x_1-x_2+u\notag\\
	y =&\ x_1.\label{eq:H_1 in example}
\end{align}
The system is a nonlinear OSNI system with the storage function
\begin{equation*}
	V(x) = \frac{1}{4}x_1^4+\frac{1}{2}x_1^2+\frac{1}{2}x_2^2,
\end{equation*}
which satisfies inequality (\ref{eq:dissipativity of OSNI}). 
According to Theorem \ref{theorem:stability}, if the required conditions are satisfied, then the system $H_1$ of the form (\ref{eq:H_1 in example}) can be asymptotically stabilized by applying the HIGS $\mc H$ of the form (\ref{eq:HIGS}) in positive feedback, as shown in Fig.~\ref{fig:closed-loop interconnection}.

Assumptions I and II are satisfied. For Assumption III, given any nonzero input $\bar u$ to the system $H_1$, suppose the HIGS output remains constant $x_h=\bar x_h$, then in the case of integrator mode, we have $e = 0$. This implies that $x_1 \equiv e \equiv 0$. Hence, $x_2 = \dot x_1 \equiv 0$ and $\dot x_2 = 0$. Therefore, $\bar u = x_1^3+x_1=0$. In this case, the system must be in the gain mode. We have $e = \bar e = \frac{1}{k_h}\bar x_h$. This implies that $x_1 = \bar x_1 = \bar e = \frac{1}{k_h}\bar x_h$. Similarly, we have $x_2\equiv 0$ and $\dot x_2 =0$, which implies that $\bar u = \bar x_1^3+\bar x_1$. Hence, we need to show that
\begin{equation*}
	\bar x_1^3+\bar x_1 \neq k_h \bar x_1.
\end{equation*}
By choosing $k_h\in (0,1]$, this inequality is satisfied except for $\bar x_1=0$. This implies that Assumption III is satisfied. 
Since $k_h \leq 1$, then the storage function of the interconnection given by
\begin{equation*}
	W = \frac{1}{4}x_1^4+\frac{1}{2}x_1^2+\frac{1}{2}x_2^2+ \frac{1}{2k_h}x_h^2-x_hx_1
\end{equation*}
is positive definite. According to Theorem \ref{theorem:stability}, the interconnection of the system $H_1$ and the HIGS of the form (\ref{eq:HIGS}) with $k_h\leq 1$ is asymptotically stable.

\begin{figure}[h!]
\centering
\psfrag{States Trajectories}{\small State Trajectories}
\psfrag{States}{\small States}
\psfrag{time (s)}{\small Time $(s)$}
\psfrag{x1}{\scriptsize$x_1$}
\psfrag{x2}{\scriptsize$x_2$}
\psfrag{xh}{\scriptsize$x_h$}
\psfrag{10}{\scriptsize$10$}
\psfrag{0}{\scriptsize$0$}
\psfrag{-10}{\hspace{-0.12cm}\scriptsize$-10$}
\psfrag{-20}{\hspace{-0.12cm}\scriptsize$-20$}
\psfrag{5}{\scriptsize$5$}
\psfrag{15}{\scriptsize$15$}
\hspace{0.5cm}\includegraphics[width=8cm]{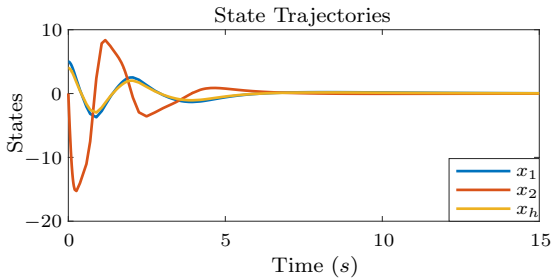}
\caption{State trajectories of the state variables of the closed-loop interconnection of the system $H_1$ given by (\ref{eq:H_1 in example}) and the HIGS $\mc H$ given by (\ref{eq:HIGS}).}
\label{fig:simulation}
\end{figure}

By choosing $k_h = 0.8$ and $\omega_h = 20$ and letting the system $H_1$ have an initial displacement $x_1 = 5$, the state trajectories of the closed-loop interconnection of the system $H_1$ and the HIGS $\mc H$ are achieved via simulation and are shown in Fig.~\ref{fig:simulation}.

\section{Conclusion}\label{sec:conclusion}
In this paper, the nonlinear NI systems theory has been extended to switched systems. We have provided definitions for switched nonlinear NI and OSNI systems. It has been obtained that the closed-loop interconnection of two switched nonlinear NI systems with external inputs and outputs is also switched nonlinear NI, and the closed-loop interconnection is Lyapunov stable under zero input. Under some assumptions, the closed-loop interconnection of a switched nonlinear NI system and a switched nonlinear OSNI system has been shown to be asymptotically stable under zero input. This stability result has been illustrated using an example where a nonlinear mass-spring-damper system is asymptotically stabilized by a HIGS controller, which is switched nonlinear NI.

\begin{ack}
The author Kanghong Shi thanks Nastaran Nikooienejad for sharing the simulation file of their previously coauthored paper \cite{shi2022negative}, which was referred to for the simulation of the example in the present paper.
\end{ack}


\end{document}